\documentclass[aps,pre,reprint,groupedaddresses]{revtex4-1}

\usepackage{amsmath}
\usepackage{amssymb}
\usepackage{graphicx}% Include figure files
\usepackage{dcolumn}% Align table columns on decimal point
\usepackage{bm}% bold math
\usepackage[separate-uncertainty=true,multi-part-units=single]{siunitx}
\usepackage{url}

\newcommand{\order}[1]{O\!\left\{ #1 \right\}}
\newcommand{\abs}[1]{\left\vert #1 \right\vert}
\renewcommand{\vec}[1]{\mathbf{#1}}
\newcommand{\Var}[1]{\mathrm{Var}\!\left\{ #1 \right\}}

\begin{document}

\preprint{APS/123-QED}

\title{Charged hydrophobic colloids at an oil/aqueous phase
interface}
\author{Colm~P.~Kelleher} \affiliation{Department of Physics and
Center for Soft Matter Research, New York University, 4 Washington Place, New York, New York 10003, USA\phantom{.}}

\author{Anna~Wang} \affiliation{Harvard John A. Paulson School of
Engineering and Applied Sciences, Harvard University, Cambridge, Massachusetts 02138, USA}

\author{Guillermo Iv\'an Guerrero-Garc\'ia } \affiliation{Department of
Materials Science and Engineering, Northwestern University, Evanston,
IL, USA, and Instituto de Fisica, Universidad Autonoma de San Luis
Potosi, San Luis Potosi, Mexico}

\author{Andrew~D.~Hollingsworth, Rodrigo E. Guerra,
Bhaskar~Jyoti~Krishnatreya, David~G.~Grier} \affiliation{Department of
Physics and Center for Soft Matter Research, New York
University, 4 Washington Place, New York, New York 10003, USA\phantom{l}}

\author{Vinothan~N.~Manoharan} \affiliation{Harvard John A. Paulson
School of Engineering and Applied Sciences, Harvard University and
Department of Physics, Harvard University, Cambridge, Massachusetts 02138, USA}

\author{Paul~M.~Chaikin} \affiliation{Department of Physics and Center
for Soft Matter Research, New York University, 4 Washington Place, New York, New York 10003, USA\phantom{i}}

\date{\today}% It is always \today, today, % but any date may be explicitly specified

\begin{abstract} 
Hydrophobic PMMA colloidal particles, when dispersed in oil with a
relatively high dielectric constant, can become highly charged. In the
presence of an interface with a conducting aqueous phase, image charge
effects lead to strong binding of colloidal particles to the
interface, even though the particles are wetted very little by the
aqueous phase. In this paper, we study both the behavior of individual
colloidal particles as they approach the interface, and the
interactions between particles that are already interfacially
bound. We demonstrate that using particles which are minimally wetted
by the aqueous phase allows us to isolate and study those interactions
which are due solely to charging of the particle surface in
oil. Finally, we show that these interactions can be understood by a
simple image-charge model in which the particle charge $q$ is the sole
fitting parameter.
\end{abstract}

\maketitle

\section{\label{sec:Intro}Introduction} 
Understanding the behavior of colloidal particles at fluid interfaces
is a long-standing \cite{RamsdenPRSL1903} and actively studied problem
in soft condensed matter physics
\cite{FrydelPRL2007,OttelDietrichLangmuir2008,CuiSceince2013,PoulichetPNAS2015}.
Extensive experimental and theoretical work has been carried out on
interactions between particles that are partially wetted by both
fluids, that is, systems where the equilibrium contact angle
$\theta_{C}$ falls in the range $0^{\circ}<\theta_{C}<180^{\circ}$. As
noted by Pieranski \cite{PieranskiPRL1980}, the presence of a fluid
interface can lead to a charge asymmetry in the vicinity of each
wetted particle, and hence to interactions which are dipolar in
form. Indeed, the $r^{-4}$ force law characteristic of dipole-dipole
repulsion has been observed in many experiments \cite{AveyardPRL2002,
ParkSoftMatter2014, ParoliniJCPM2015}.

However, various aspects of the interactions between interfacial
particles are still not well-understood
\cite{McGortyMatToday2010}. For instance, interfacial colloids may
form repulsive crystals or fractal aggregates
\cite{ReynaertLangmuir2006}, or may self-assemble into more complex
mesoscopic structures \cite{GhezziJPhysCondMat1997}. The interactions
responsible for this collective behavior are typically very sensitive
to the protocol used to prepare the samples
\cite{ParkFurstSoftMatter2011,GaoNature2014}, are highly non-uniform
\cite{ParkSoftMatter2010}, and are strongly time-dependent
\cite{GaoNature2014}.

\begin{figure}[t!]
\begin{centering}
\includegraphics[scale=0.44]{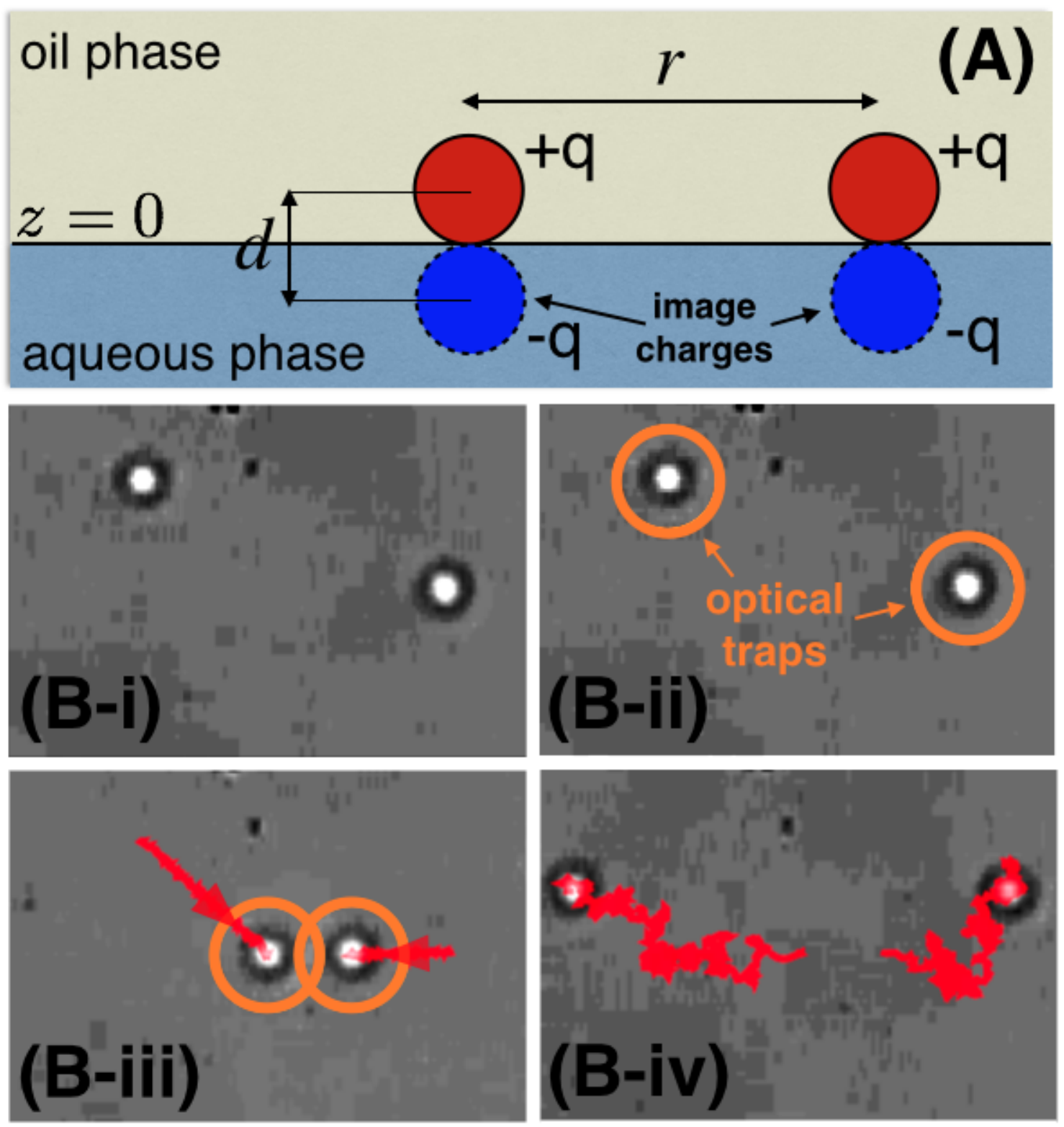}
  \caption{(Color online) (A) Once bound to the interface, non-wetting
colloids repel electrostatically. (B) Optical
micrographs of a single run of an experiment to probe the repulsive
force between pairs of interfacial colloids, each with a diameter $d =
\SI{1.1}{\um}$.  (i) \& (ii) Two interfacial particles, more than
\SI{20}{\um} away from any others, are identified using a particle
tracking algorithm, and (iii) automatically dragged to a
pre-assigned location using laser tweezers. (iv) The particles are
released from the tweezers, and move freely along the interface. Their
trajectores (shown in red) are recorded throughout.}
    \label{fig:ImageCharge}
\end{centering}
\end{figure} 

To explain these complicated interactions, different authors have
proposed various modifications or extensions of Pieranski's simple model.
These include mechanisms for interparticle attraction, such as from
inhomogenous charge distribution on the particle surface
\cite{ChenPRE2009}, and interparticle repulsion, for example by
charging of the particle surface in oil \cite{AveyardLangmuir1999,
GaoNature2014}. Finite-ion-size effects in the aqueous phase have been
proposed to explain the anomalous dependence of the interparticle
force on salt concentration \cite{Masschaele2010}, while irregular
pinning of the contact line on the colloid surface introduces
anisotropic capillary forces between particles \cite{StamouPRL2000,
KazNatMat2011}. Moreover, since all these effects can in principle
occur at the same time in the same sample, it is difficult to
disentangle them.

In this paper, we report measurements of the interactions between
colloidal spheres at an oil/aqueous phase interface in a system with
two useful properties.  First, the spheres are embedded almost
entirely in the oil phase and are wetted very little, or not at all,
by the aqueous phase.  Second, the oil has a dielectric constant which
is large compared to that of typical hydrocarbon oils, and so can
harbor mobile charges. These properties allow us to isolate and
explore how the interparticle interactions are influenced by
electrostatic charges on the particles' surfaces.  Similar systems
have been studied previously \cite{Leunissen2007}, particularly for
the insights they offer into the proliferation and dynamics of
topological defects in two-dimensional curved spaces
\cite{IrvineNature2010}.  By elucidating the nature of the
interactions in this system, we also hope to cast new light on these
phenomena.

We study two different aspects of the behavior of colloids in this
system: the approach and binding of individual particles to the
oil/aqueous phase interface, and the repulsive force between
interfacially bound colloids. We show that both sets of observations
can be quantitatively described by a simple electrostatic model in
which the aqueous phase plays the role of a conducting substrate, and
the particle charge $q$ is the only adjustable parameter. This model
is shown schematically in Fig. \ref{fig:ImageCharge}(A).

\section{Materials}
Our experimental system is composed of poly(methyl methacrylate)
(PMMA) spheres, dispersed in oil, in the vicinity of a glycerol/water
mixture (``the aqueous phase''). 

\subsubsection{Preparation of Glassware \& Sample
Chambers \label{cleaning}} The glass we use to store the particles
and to construct sample chambers is sonicated for \SI{10}{\second}
in \SI{5}{wt.\percent} Contrad 70 detergent, followed by sequential
rinsing in de-ionized water, acetone and isopropanol. The glass is
then blown dry with an N$_2$ sprayer and placed in an oven at \SI{70}{\celsius} for at
least \SI{15}{\minute} prior to use. We note the following
exception: the sample chamber we use in the experiment described in
Section \ref{gofr} consists of a glass capillary tube of internal
dimensions \SI{100}{\micro\meter} $\times$ \SI{2}{\milli\meter} $\times$
\SI{5}{\centi\meter} (VitroTubes) which is
ultrasonicated in Millipore water for \SI{10}{\second}, and finally
dried in an oven at \SI{70}{\celsius} for \SI{2}{\hour}.  Where
necessary, we use a glycerol buffer phase to ensure that the oil never
comes into contact with the Norland optical adhesive we
use to seal the samples.

\subsubsection{Fluid Phases} 
The aqueous phase consists of
\SI{10}{mM} NaCl in a \SI{70}{wt.\percent} glycerol solution, while
the oil phase consists of a 5:3:2 v/v mixture of cyclohexyl bromide
(CHB), hexane and dodecane. To prevent ionic contamination of the oil
phase, we filter and store it according to the protocols
described in Refs. \cite{MirjamLeunissenPhD} and
\cite{Leunissen2007}.  Using the formula given in
\cite{YoolengaPhysica1965}, we estimate that this oil has a relative
dielectric constant $\epsilon_{r} = \num{4.2} $, which is much lower
than water ($\epsilon_{r} \approx \num{80}$), but significantly higher
than alkanes such as decane ($\epsilon_{r} \approx \num{2}
$). Theoretical estimates \cite{ZwanikkenJPhys2008} indicate that an
oil with $\epsilon_{r} = \num{4}$, in contact with a water
reservoir, will reach an equilibrium ionic concentration with a Debye
screening length $\lambda_{D}$ of approximately \SI{50}{\um}, which is
far greater than the length scales probed in our experiments.

\begin{figure}[t!]
\center
\includegraphics[scale=0.5]{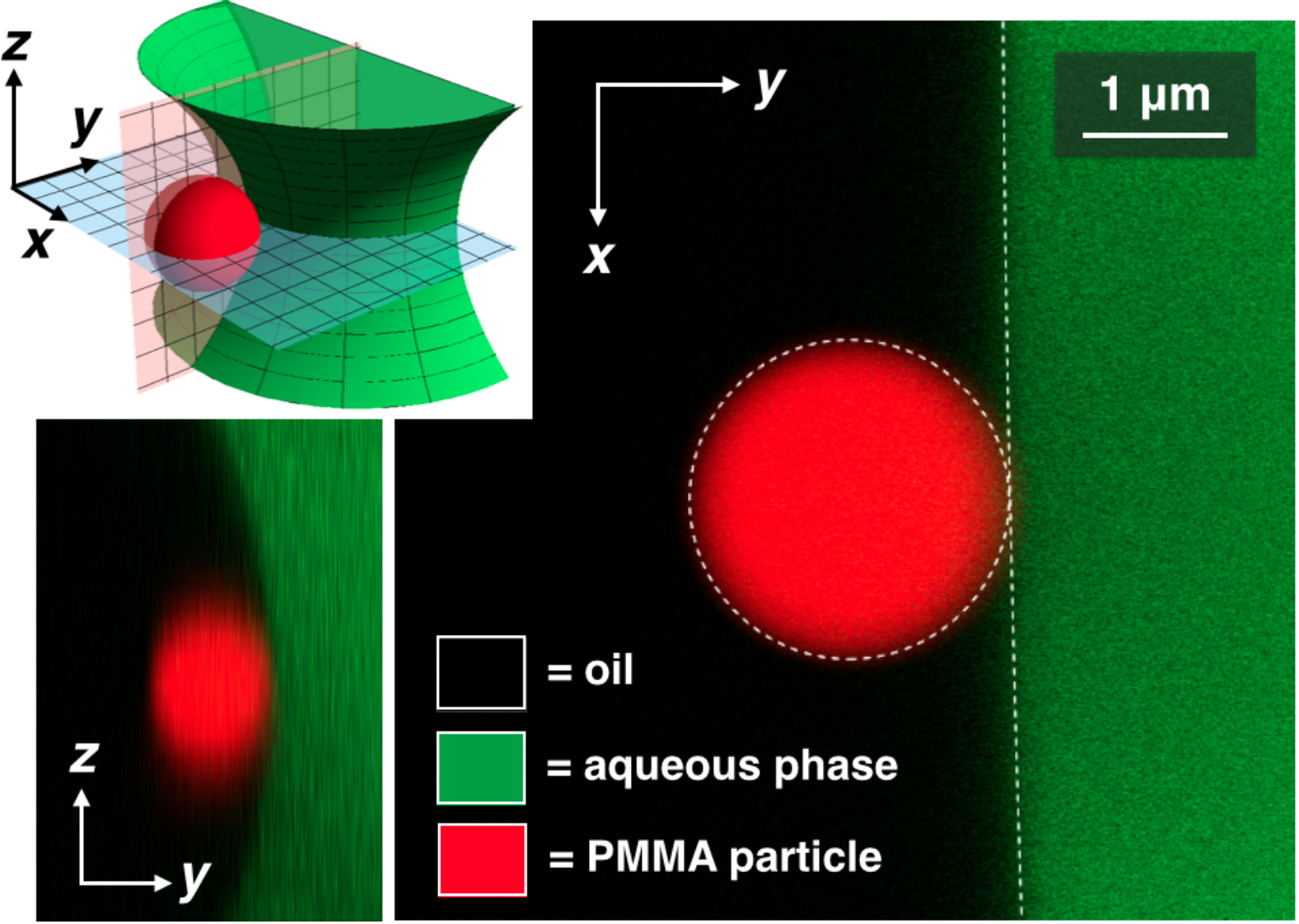}
  \caption{(Color online) Horizontal and vertical slices of a confocal
micrograph of a single PMMA sphere electrostatically bound to the neck
of a capillary bridge droplet.  Top Left: Schematic of the geometry of
the particle and interface. The curvature of the capillary bridge is
exaggerated for clarity.  Main Figure: The dashed white curves show
the result of an edge-finding routine designed to measure the contact
angle. For this particle, $\theta_C =
\SI{180}{\degree}^{+\SI{0}{\degree}}_{-\SI{9}{\degree}}$.}
  \label{fig:NonWetting} 
\end{figure}

\begin{figure*}[t!]
\includegraphics[scale=0.55]{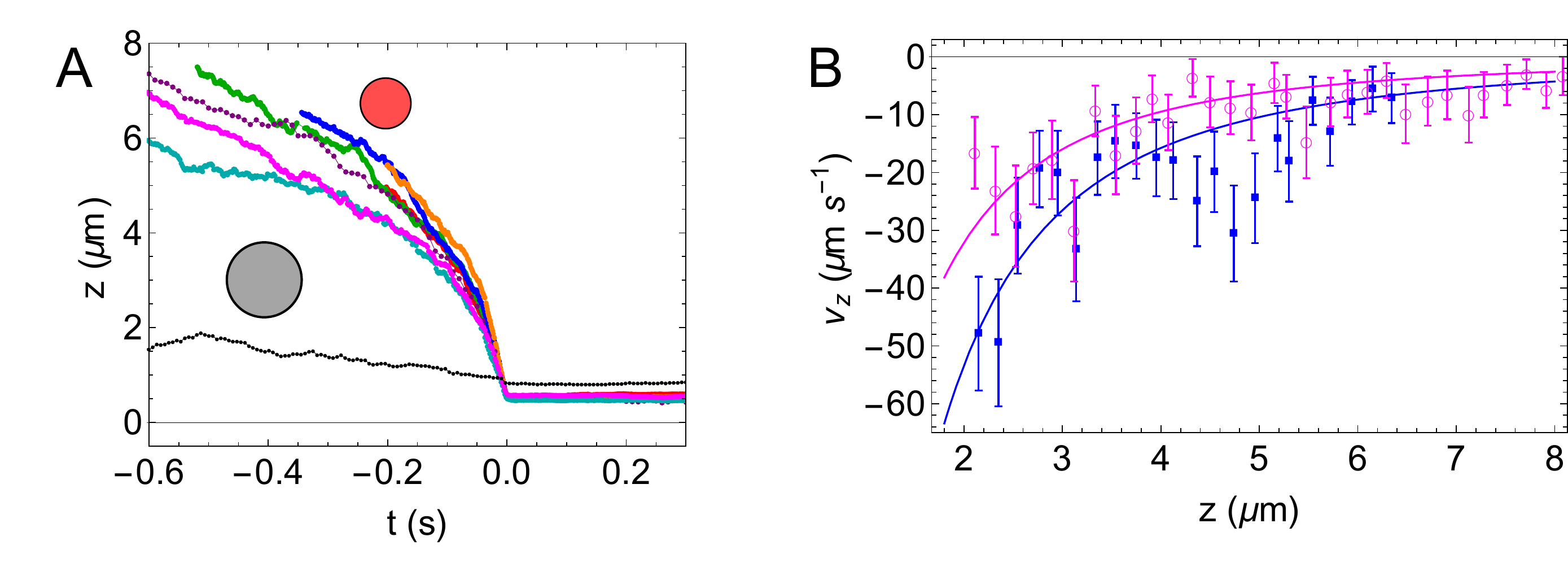}
\caption{(Color online) (A) The colored trajectories are obtained by digital
holographic microscopy of seven different $d = \SI{1.1}{\um}$
colloidal particles approaching the oil/aqueous phase interface. For
each trajectory, we have set $t = 0$ as the time of the attachment
event, and $z = 0$ as the position of the fluid interface.  The
charged particles are strongly attracted to the interface, achieving
speeds of up to \SI{80}{\um\per\second}. For comparison, these
particles have a sedimentation velocity of \SI{0.04}{\um\per\second}.
We have also plotted (in black) the trajectory of a $d =
\SI{1.6}{\um}$ PMMA particle in a system where the oil phase consists
of pure decane \cite{AnnasPaper}. In the absence of CHB, the particle
is far less charged, and approaches the interface much more slowly
than in the presence of CHB.  The red and gray disks indicate the
sizes of the colloidal particles, to scale on the $z$-axis. (B) Plot
of the velocities of two of the trajectories, as a function of
distance from the interface. The error bars indiate the expected width
of the velocity distribution, calculated from the bulk diffusion constant
$D_{0}$. For each data set, the best fits of the model given by
Eqs.~\eqref{eq:BindingForce}, \eqref{eq:Fv} and \eqref{eq:zDrag} are
also shown.}
  \label{fig:Ztrajectories}
\end{figure*}

\subsubsection{Colloidal Particles} The PMMA microparticles are
sterically stabilized with covalently bound poly(12-hydroxystearic
acid) \cite{ElsesserHollingsworth2010}.  Such particles have a surface
charge that might be caused by adsorption of positively charged
species resulting from the decomposition of CHB
\cite{MirjamLeunissenPhD}, chemical coupling of an amine catalyst
during particle synthesis \cite{vanderLindenLangmuir2014}, or some
combination of these mechanisms. In some of our experiments, we use
spheres that are fluorescently labeled with absorbed rhodamine 6G dye
\cite{ElsesserEtAl2011}.  We find that dyeing the particles does not
affect their measured interactions. 

\section{Measurement of the Contact Angle $\theta_{C}$}
To verify that our particles remain entirely immersed in the oil and
are not wetted by the aqueous phase, we measure their contact angle
directly by fluorescent confocal microscopy. We do this by preparing a
low-concentration dispersion of PMMA particles of mean diameter \SI{2.6}{\um} 
in oil, and flow the dispersion into a channel containing several
capillary bridges of the aqueous phase. To create these capillary
bridges, we first use a sprayer to deposit droplets (of typical
diameter \SIrange{10}{100}{\um}) of the aqueous phase on a cover
slip. We then place the cover slip, droplet side down, on Dura-lar
spacers of thickness \SI{25}{\um} which have been placed on a
microscope slide. The larger droplets come into contact with the
microscope slide, and spontaneously form capillary bridges. We use a
two-channel Leica TCS SP5 II confocal microscope, with a $63\times$ NA
1.4 oil-immersion objective lens to simultaneously image the particles
and the aqueous phase. For these studies, the aqueous phase is
fluorescently labeled by replacing some or all of the NaCl with
fluorescein sodium salt. Since the oil and aqueous phases are
refractive-index matched to within approximately 1\%, optical
artifacts arising from the curvature of the interface are minimized.

As shown in Fig.~\ref{fig:NonWetting}, some of the particles bind to
the neck of a capillary bridge, presumably by electrostatic
forces. Using inbuilt edge-detection algorithms from the
commercial software package \textit{Mathematica}, we identify the
edges of the capillary bridge and the colloidal particle.  The contact
angle is calculated from these data.  A typical confocal slice,
overlaid with the results of the edge-finding routine, is shown in
Fig.~\ref{fig:NonWetting}. For our system, we measure the best-fit
contact angle of the particles to be in the range
\SIrange{171}{180}{\degree}, consistent with the results of
Ref.~\cite{Leunissen2007}.  All measurements are consistent with a
contact angle of \SI{180}{\degree}.

\section{Colloid-Interface Interaction \label{sec:CI}} 
We probe the interaction of individual colloidal spheres with a flat,
horizontal oil/aqueous phase interface by measuring their trajectories
as they move through the oil phase toward the interface. Because the
particles move at speeds up to \SI{80}{\um\per\second}, too fast to
track with confocal microscopy, we measure their trajectories with
digital holographic microscopy. In this technique, an incident
monochromatic plane wave scatters from a spherical colloidal
particle. Using the apparatus described in Refs.~\cite{AnnasPaper}
and~\cite{KazNatMat2011}, we digitally record the image that results
from interference of the scattered light with the incident plane wave
\cite{GaborPRS1949,SchnarsJuptnerMST2002}. Fitting the interference
pattern predicted by Lorentz-Mie theory to the recorded hologram
\cite{LeeOE2007} gives the particle's three-dimensional position with
\SI{3}{\nm} precision over a \SI{50 x 50 x 50}{\um} volume at time
intervals as low as \SI{1}{\ms}.  Since the oil
and aqueous phases have well-matched refractive indices, we fit the
data using functions appropriate for scattering from a dielectric
sphere immersed in a medium of uniform refractive index
\cite{HolopyWebsite}. To avoid interference from multiple particles in
the same image, PMMA-in-oil dispersions are prepared at volume
fractions below \num{e-6}. As well as position data, the holographic
measurements yield estimates for the diameters of the particles,
with nanometer precision, and their refractive indices, which can be
used for consistency checks. The colloidal particles we use for these
and all subsequent studies have a mean diameter $d = \SI{1.08}{\um}$ and polydispersity
5\%.

To understand the observed trajectories of the particles as they
approach the interface (Fig.~\ref{fig:Ztrajectories}), we construct an
equation of motion involving the electrostatic force and drag. Because
the aqueous phase contains dissolved salt ions that act as free
charges, we treat it as a good conductor. A sphere of
charge $q$ whose center is at height $z$ above a flat conducting surface is
attracted towards its image charge with a force \cite{jacksonEandM}
\begin{equation}
  \label{eq:BindingForce} 
  F_z(z) 
  =
  -\frac{q^2}{16 \pi \epsilon_r \epsilon_0 z^2}.
\end{equation} 

Because the motion is overdamped, the speed with which the sphere
approaches the interface is given by
\begin{equation}
  \label{eq:Fv} 
  v_z(z) = \frac{ F_z(z)}{\gamma_\perp(z)},
\end{equation} 
where $\gamma_\perp(z)$ is the viscous drag coefficient for motions
perpendicular to the fluid-fluid interface located in the plane $z = 0$.  Lee
and Leal \cite{Lee1979} find that
\begin{equation}
  \label{eq:zDrag} 
  \frac{\gamma_{\perp}(z)}{\gamma_{0}}
  = 
  1
  + 
  \frac{3}{16} \frac{d}{z} \frac{2+3\lambda}{1+\lambda} 
  +
  \left(
    \frac{3}{16} \frac{d}{z} 
    \frac{2+3\lambda}{1+\lambda}
  \right)^2
  + 
  \order{\frac{d^3}{z^3}},
\end{equation} 
where $\lambda = \mu_{\text{aq}}/\mu_{\text{oil}}$ is the ratio of
dynamic viscosities of the two fluid phases and $\gamma_0 = 3 \pi
\mu_{\text{oil}} d$ is the Stokes drag on a sphere far from any
boundaries.  For our system, $\mu_\text{aq} =
\SI{23}{\milli\pascal\second}$ (from tabulated values) and
$\mu_\text{oil} = \SI{1.7}{\milli\pascal\second}$ (from our
measurements of the particle diameter and diffusion constant in bulk
oil) so that $\lambda = \num{14}$.  We estimate $D_{0}$, the 3D
diffusion constant of the particles far from any boundaries, by
measuring $D_\parallel(z)$, the diffusion coefficient parallel to the
interface. We calculate $D_\parallel(z)$ from those parts of the
trajectories where $z > \SI{2}{\um}$. From the hydrodynamic theory in
Ref.~\cite{Lee1979}, we estimate that the error in approximating
$D_{0}$ by $D_\parallel(z > \SI{2}{\um})$ is around
\SI{6}{\percent}. We then use the fluctuation-dissipation relation,
$\gamma_0 = k_{B}T/D_{0}$, to obtain the drag coefficient at the
absolute temperature $T = \SI{293}{\kelvin}$.  A typical value for the
spheres in this study is $\gamma _0 =$ \SI{17}{\nano \newton \second
\per \meter}.

Applying the model consisting of Eqs.~\eqref{eq:BindingForce},
\eqref{eq:Fv} and \eqref{eq:zDrag} to the data in
Fig.~\ref{fig:Ztrajectories} yields good agreement with the measured
velocities for a mean sphere charge $q = \SI{530(30)}{e}$, where e is
the elementary charge, and the uncertainty is given by the standard
error of the mean.

\textit{A priori}, we cannot exclude the possibility of the presence of
significant amounts of (positive or negative) surface charge $\sigma$ on the oil/aqueous
phase interface \cite{Leunissen2007}, which would require including an
extra force $q \sigma / \epsilon_r \epsilon_0$ on the right-hand side
of Eq.~\eqref{eq:BindingForce}. Treating $\sigma$ as a fit parameter in
this expanded model yields as an upper bound $\abs{\sigma} <
\SI{0.2}{e \per\square\um}$, but neither improves the quality of the
fits nor significantly affects our estimate of the particle charge
$q$. We therefore omit $\sigma$ from the model. We also neglect the
gravitational force because, over the measured range of $z$, it is
negligible compared to the image-charge interaction:
$\abs{F_{\text{grav}}} < 0.02 \abs{F_z}$.

These measurements establish that the force drawing the PMMA spheres
to the interface is consistent with image-charge attraction and
provide an estimate of the single-sphere charge. We next investigate
how that charge influences the interaction between spheres at the
interface.

\begin{figure}[t!]
\includegraphics[scale=0.5]{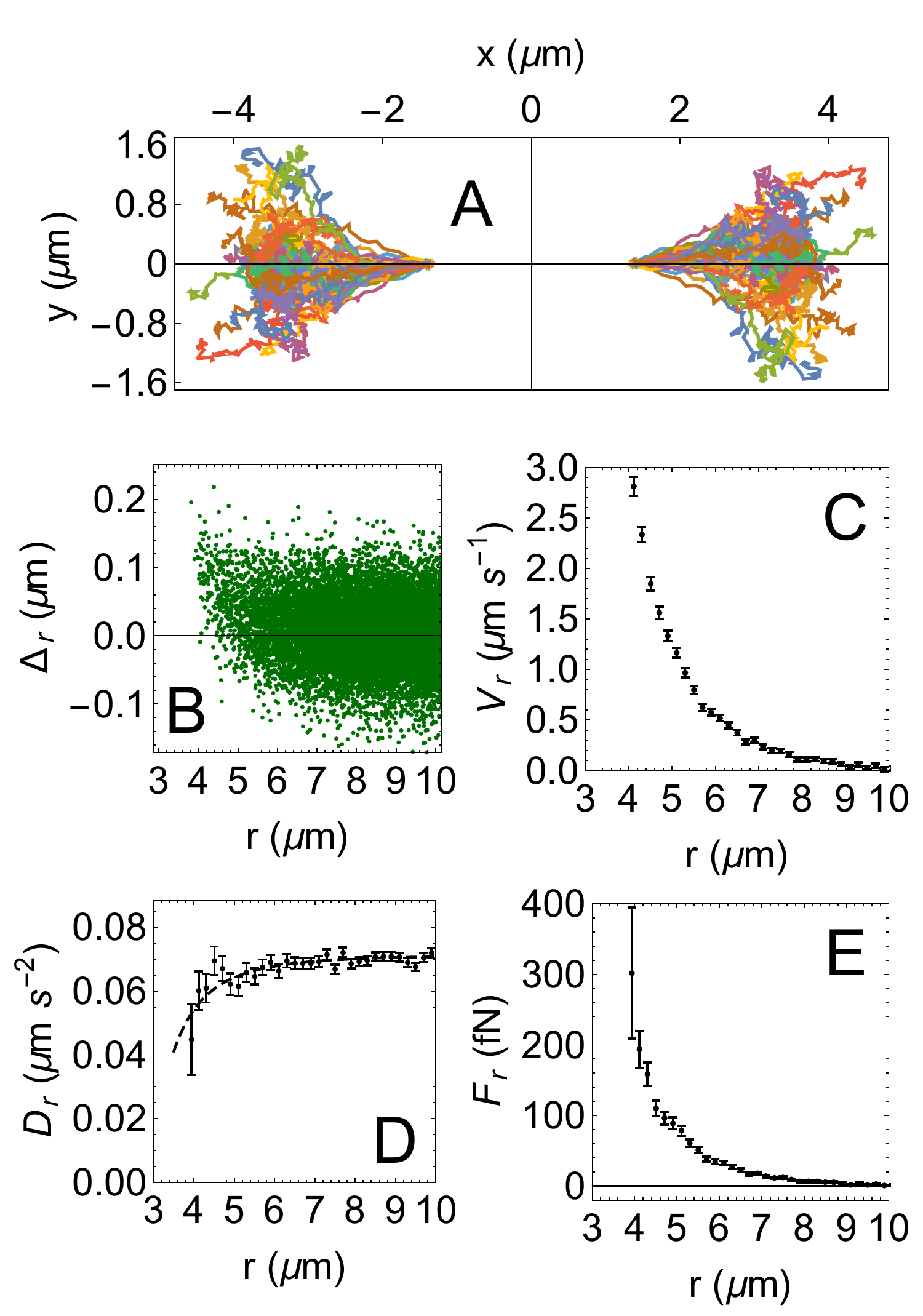}
\caption{\label{fig:TweezerExpt} (Color online) Stages in the analysis of the data
from the catch-and-release experiment to probe the repulsive force
between a specific pair of interfacial colloids.  (A) Overlay of
180 post-release trajectories of a single pair of colloidal
particles. At each instant, the positions are plotted in the
center-of-mass frame.  (B) Frame-to-frame radial displacement,
plotted as a function of center-to-center separation $r$.  (C)
Radial velocity, obtained by binning and averaging the data in
B. (D) The radial diffusion constant $D_{r}(r)$ is obtained from
the variance of data in each bin. The dashed line is a guide to the
eye, and highlights the $r$-dependence of $D_{r}$, which we attribute
to hydrodynamic interaction between the particles. We obtain the
radial drag coefficient $\gamma_{r}(r)$ by using the
fluctuation-dissipation relation, $\gamma_{r}(r) = k_{B}T/ D_{r}(r)$.
(E) The radial force is found by multiplying the velocity by the
drag coefficient, $F_{r}(r) = \gamma_{r}(r) V_{r}(r)$.}
\end{figure}

\section{Pair Interaction of Interfacial Colloids} 
In this section, we describe the results of three independent experiments
for measuring the force between interfacially bound colloidal
particles as a function of interparticle separation $r$.  The results
are all consistent with an electrostatic model in which the charge $q$ on
a single sphere is $\SI{570(30)}{e}$, which is in turn consistent
with the result described in the previous section.

We treat the colloidal particles as spheres of uniform surface charge
sitting directly above the aqueous phase, which, as in the previous
section, plays the role of a conducting substrate. As shown in
Fig.~\ref{fig:ImageCharge}(A), pairs of spheres are repelled by each
other's charges, but are attracted to their neighbors' image charges.
All our measurements take place in the regime where interparticle
separations are large compared to the colloid diameter, but
small compared to the Debye length in the oil phase, so that $d^{2}
\ll r^{2} \ll \lambda_{D}^{2}$. In this limit, the net interaction force,
$F_r(r)$, between pairs of spheres with center-to-center separation
$r$ takes a dipolar form,
\begin{equation} F_{r}(r) = -\frac{q^{2}}{4\pi \epsilon_{r}
\epsilon_{0}} \frac{\text{d}}{{\text{dr}}}\Bigg{(} \frac{1}{r} -
\frac{1}{(r^{2}+d^{2})^{1/2}} \Bigg{)} \simeq
\frac{3B}{r^{4}} 
\label{eq:Fr},
\end{equation} 
where the force constant $3B$ is related to the particle charge by $B
= q^{2}d^{2}/8 \pi \epsilon_{r} \epsilon_{0}$.

\subsection{``Catch-and-Release'' Laser Tweezer Experiments} 
Our first measurement of the repulsive force between a pair of
interfacial particles proceeds by forcing the particles close together
with a pair of optical tweezers and then releasing them. We record and
analyze the resulting trajectories to find the interparticle force, as
illustrated in Fig.~\ref{fig:TweezerExpt}. 

For these experiments, and also those of Section
\ref{sec:CrystalElasticity}, we prepare samples that contain many
small (diameter \SIrange{100}{500}{\micro\meter}), almost-flat interfaces
that are isolated from each other. To make these interfaces, a cover
slip is immeresed in a bath of KOH-saturated isopropanol for
\SI{1}{\hour} prior to undergoing the treatment described in Section
\ref{cleaning}. We use a sprayer to deposit droplets of the aqueous
phase onto the cover slip, which is then incorporated into the
construction of a capillary channel. Finally, the channel is filled
with the particle dispersion and sealed with Norland optical
adhesive. Following this protocol, each droplet of the aqueous phase
forms a roughly spherical cap on the glass surface, with a contact
angle of \SI{1}{\degree} or less. The resulting interface is flat
enough to allow bright-field imaging of interfacial particles, which
adsorb to the interface because of the electrostatic attraction
described in Section \ref{sec:CI}. Effectively random factors, such as
how far a given droplet is from the entrance of the capillary channel,
influence how many particles are deposited on each interface. Thus,
within a in a single sample cell, we obtain many isolated interfaces,
each with a different interfacial density.

To measure the interparticle force, we first identify an interface at
sufficently low particle density that only two spheres are in the field of
view of the microscope. A particle tracking algorithm then locates the spheres
\cite{CrockerGrier1994,CrockerGrier1996}. Once located, the spheres
are confined in holographic optical traps projected at their position
(``catch'') \cite{Curtis2002, DufresnseGrierRSI1998, PolinOE2005}.
The holographic trapping system is created with a \SI{1064}{\nm} laser
(IPG Photonics YLR-10-1064-LP) whose wavefronts are modified using a
computer-controlled liquid-crystal spatial light modulator (Holoeye
Pluto). The resulting light pattern is relayed to an objective lens
(Nikon Plan Apo, NA 1.45 $100\times$, oil immersion) that focuses the
traps into the sample.  The traps drag the particles towards one
another until they reach a pre-assigned minimum distance, at which
point the traps are instantaneously displaced tens of microns in the
direction perpendicular to the imaging plane, allowing the particles
to move freely along the interface (``release''). The trajectories of
the particles are recorded by video microscopy, and the coordinates of
the centers of the particles, $\mathbf{r}_{1}$ and $\mathbf{r}_{2}$,
are measured using publicly available tracking routines
\cite{CrockerGrier1996, CrockerWeeksTracking}.  This procedure, shown
in Fig.~\ref{fig:ImageCharge}(B), is fully automated, and, for a given
pair of particles, is repeated as many as several hundred times.

\begin{figure}[h!]  \center
\includegraphics[scale=0.5]{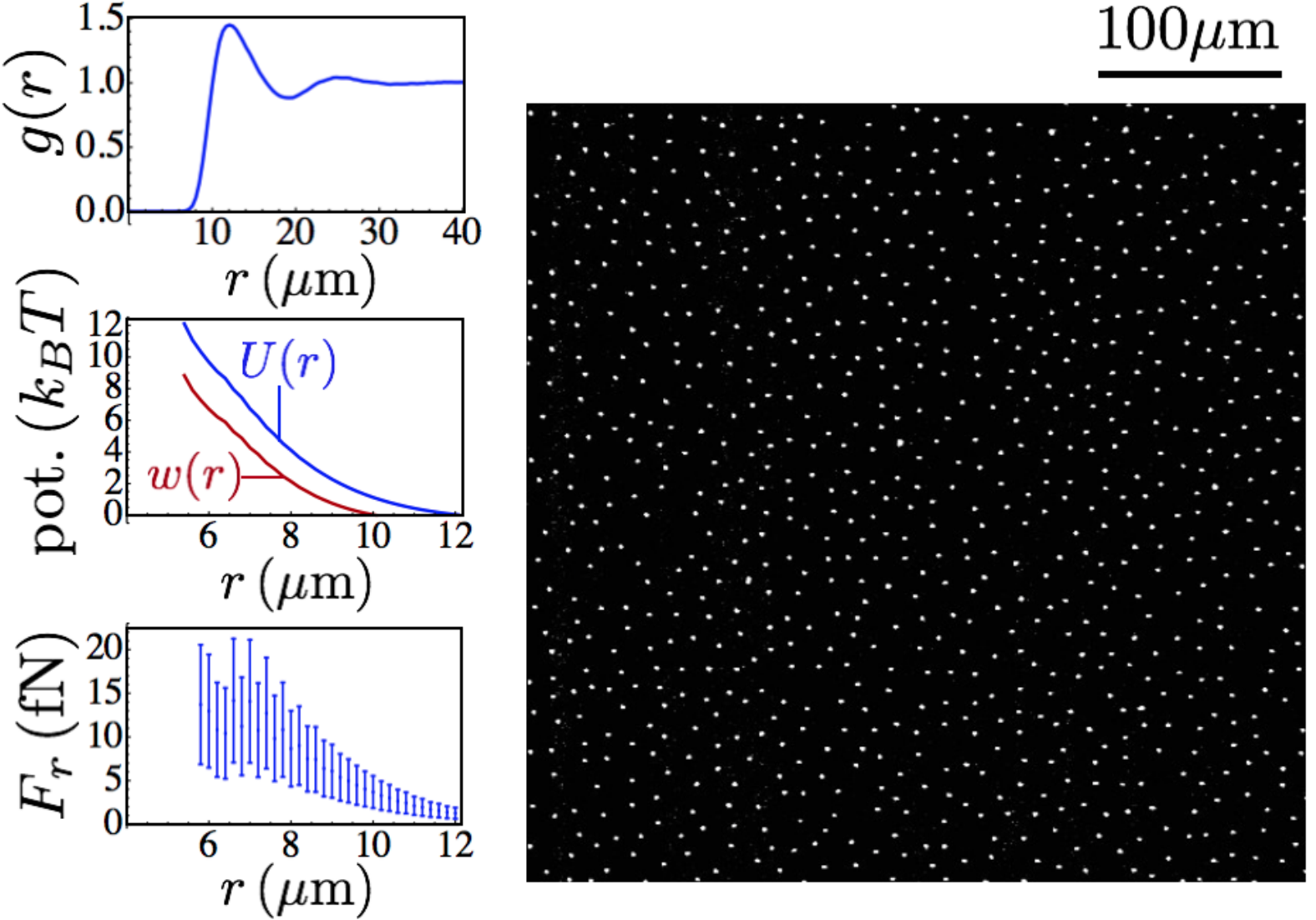}
\caption{\label{fig:GofR} (Color online) Main Image: Confocal
micrograph of PMMA particles bound to a glycerol/water interface. This
image is a single snapshot taken from a $\SI{1}{\hour}$ movie, with an
average particle density of 0.005$\,\mu$m$^{-2}$. Left
Column: A series of plots showing: (i) $g(r)$, from all frames of
the movie; (ii) the potential of mean force $w(r) = -k_{B}T \log{g}$
and the pair potential $U(r)$. We use the Ornstein-Zernike equation
to obtain $U(r)$ from $g(r)$; (iii) the radial force, obtained from
$U(r)$ by numerical differentiation.}
\end{figure}

\begin{figure*}[t!]
\includegraphics[scale=0.5]{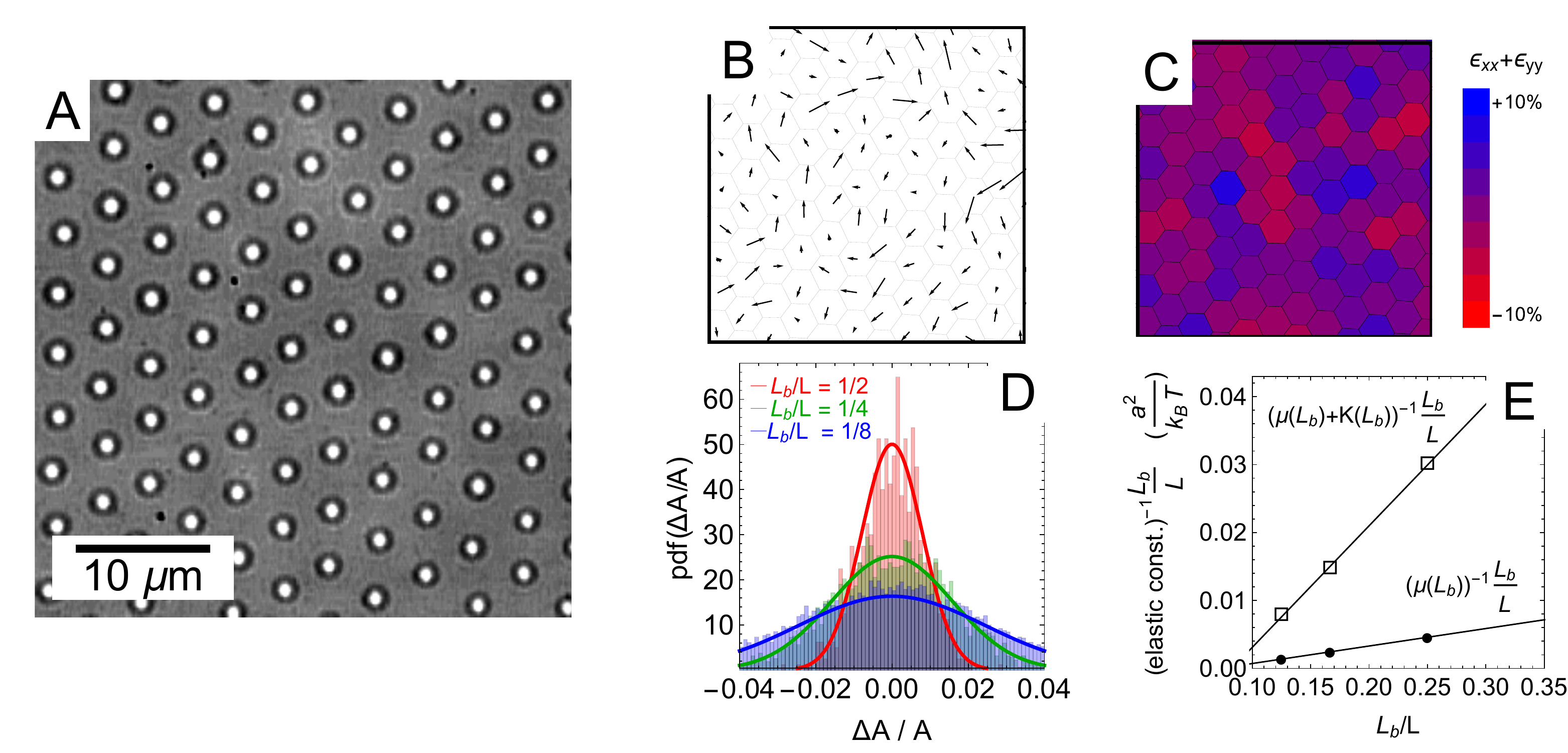}
  \caption{(Color online) Finding the elastic moduli of a 2D colloidal crystal from
video microscopy.  (A) Snapshot taken from a movie of \num{100}
statistically independent frames. The lattice constant $a$ is
$\SI{4.3}{\um}$. We analyze the fluctuations within a square box of
side length $L = 8a = \SI{34}{\um}$.  (B) The instantaneous
displacement field $\vec{u}$ corresponding to the particle
configuration shown in panel A. For clarity, the magnitude of the
displacements is exaggerated by a factor of 5. Also shown is the Voronoi
tesselation, computed from the average particle positions.  (C)
Trace of the strain tensor, $\epsilon_{xx}(\vec{r},t) +
\epsilon_{yy}(\vec{r},t)$, calculated from the displacement field
shown in panel B.  (D) Probability distribution function of the
dilatation strain $(\Delta A/A)$ for three different sub-region sizes
$L_b/L$.  Solid curves are Gaussian fits to the data.  (E) We
obtain elastic moduli of $K = 40~k_{B}T \si{\per\square\um}$ and $\mu
= 6~k_{B}T \si{\per\square\um}$, which, via
Eqs.~\eqref{eq:dipoleCrystalFormulas}, give $\Gamma = \num{305(35)}$.
Thus, for an interparticle separation of \SI{4.3}{\um}, the
interparticle force is $F_{r} = \SI{110(30)}{\femto\newton}$.}
  \label{fig:elasticConsts} 
\end{figure*}

Once a particle is released from the optical traps, its motion results
from a combination of interaction with the other particle, and
diffusion. For two subsequent frames, the displacement of the
particles along the radial direction in the center-of-mass reference
frame is
\begin{equation} \Delta_{r}(t) =\left[\vec{r}(t + \tau) - \vec{r}(t)
\right] \cdot \mathbf{\hat{r}}(t),
\end{equation} where $\vec{r}(t) $ is the instantaneous separation at
time $t$, $\vec{r}(t) = \vec{r}_2(t) - \vec{r}_1(t)$, and $\tau =
\SI{16.7}{\ms}$ is the time interval between video fields.  The
relative velocity, $v_{r}(r) = \Delta_r/\tau$, as well as the
separation-dependent drag coefficient in the radial direction
$\gamma_{r}(r)$, is found by combining the data from multiple
releases, and plotting $\Delta_{r}$ as a function of interparticle
separation $r$. The data are divided into bins along the $r$-axis,
as shown in Fig.~\ref{fig:TweezerExpt}, and the interparticle force is obtained
by using the overdamped equation of motion, $F_{r}(r) = \gamma_{r}(r)
v_{r}(r)$ \cite{CrockerGrier1994,
Crocker1997, Sainis2007}. The results of 10 such experiments, on different pairs of
particles, are shown in Fig.~\ref{fig:FinalResult}(A). Fitting
Eq.~\eqref{eq:Fr} to this data gives $q = \SI{580(30)}{e}$, which is
consistent with the value obtained from the colloid-interface
experiments in Section~\ref{sec:CI}.

\subsection{Pair Correlation Function Experiments \label{gofr}} 
In a second experiment to measure the pair interaction, we measure the
pair correlation function $g(r)$ of a system of interfacial particles
at low areal density $\rho$. We use the Ornstein-Zernike equation from
liquid state theory, with the hypernetted chain approximation, to
obtain the pair potential $U(r)$ from $g(r)$
\cite{BehrensGrier2001,HansenMcDonald2006}, and finally calculate
$F_{r}(r)$ by numerical differentiation. To sample $g(r)$ at the low
densities this method requires, we first half-fill a capillary tube
with the aqueous phase. We then place the filled end of the capillary
tube into a sample vial containing the particle dispersion.  As
the oil flows into the tube, thin patches of the aqueous phase are
left behind on the top and bottom glass surfaces. These patches, held in
place by pinning of the contact line, are typically
\SIrange{0}{2}{\micro\meter} thick, and millimeters in diameter.  We
are thus able to image regions of uniform density as large as
\SI{0.3}{\square\mm}, and which contain hundreds of particles.  We
use a Leica TCS SP5 II confocal microscope, mounted with a $10\times$
air objective lens, to collect movies which are typically
\SIrange{1}{2}{\hour} in length. Using publicly available software
\cite{CrockerWeeksTracking}, we find the positions of the particles in
each frame and obtain $g(r)$ for each movie.  Fig.~\ref{fig:GofR}
shows a snapshot of a typical sample, together with the stages of the
anaysis.

To estimate the error in finding the force in this manner, we perform
a series of Monte-Carlo simulations of point particles interacting via
a set of known interaction potentials. We choose the interaction
potentials and densities in the simulations so that they produce pair
correlation functions similar to those observed in the
experiments. From each simulation, we calculate $g(r)$, and then apply
the Ornstein-Zernike method described above to obtain $U(r)$ and
$F_{r}(r)$. We compare the $F_{r}(r)$ obtained from $g(r)$ to the
$F_{r}(r)$ curve calculated directly from the potential that we use in
the simulation. The error $\Delta F_{r}$ is given by the difference
between these two values. Since the fractional error $\Delta F_{r}/
F_{r}$ does not depend strongly on $r$, $\rho$, or the parameters
describing the interaction potential, we take it to be constant,
$\Delta F_{r}/ F_{r} = 0.5$. This value is assumed when plotting error
bars such as those shown in Fig.~\ref{fig:GofR}.

Fig.~\ref{fig:FinalResult}(B) shows the results of applying the
Ornstein-Zernike inversion procedure to samples of interfacial
colloids at two different areal densities. For each sample, we repeat
the measurement one, two and four days after preparation to confirm
that the results for $g(r)$ are time-independent, and thus reflect
equilibrium properties. Fitting this data to Eq.~\eqref{eq:Fr} gives $q =
\SI{540(30)}{e}$, which is consistent both with the value obtained
from the colloid-interface interaction experiments in Section
\ref{sec:CI} and with other pair interaction experiments described in
this section.

\subsection{Crystal Elasticity Experiments}
\label{sec:CrystalElasticity} 
Our third approach for measuring the pair interaction takes advantage
of the fact that, when confined to an interface at sufficiently high
areal density, colloidal monolayers form a hexagonally-ordered solid
phase, which is stable over time-scales of many weeks \footnote{The
limiting factor in the lifetimes of these crystals is depinning of the
boundaries of the fluid interfaces on which they sit.}. This
colloidal solid is soft enough that thermal fluctuations of the
particle positions can be measured using video or confocal
microscopy. From the resulting trajectories, we can estimate the
crystal's bulk modulus, $K$, and shear modulus, $\mu$.  These elastic
constants are related to the crystal's interaction parameter, $\Gamma$, which yields
the pair potential at the mean interparticle separation $a$
\cite{WeissJChemPhys1998,ZahnPRL2003}.  Unlike the previously
discussed measurements, which yield functional forms for the
separation-dependent interaction, interaction measurements based on
lattice elasticity require us to assume a functional form. However,
since this measurement takes place at high areal density, it can
confirm the pair-wise additivity of the interactions.

To determine the interparticle force and the elastic constants, we
first measure the instantaneous strain and rotation. For a
displacement field $\vec{u}(\vec{r},t)$ at position $\vec{r}$ and time
$t$, the instantaneous strain and rotation tensors are defined as
\begin{align*}
  \epsilon_{ij}(\vec{r},t) 
  & =
    \frac{1}{2} \left[
    \partial_i u_j(\vec{r},t) 
    + 
    \partial_j u_i(\vec{r},t)
    \right] \quad \text{and} \\
  \theta_{ij}(\vec{r},t)
  & =
     \frac{1}{2} \left[
    \partial_i u_j(\vec{r},t) 
    -
    \partial_j u_i(\vec{r},t)
    \right],
\end{align*}
respectively, where $i,j \in \{x,y\}$. Adapting these definitions to
include a displacement field $\vec{u}$ defined at a discrete set of
lattice points and times, we use the lattice calculus methods
described in the Appendix to calculate the strain and
rotation tensors from the measured set of particle positions.

For a region of area $A$, the dilatation strain is given by
\begin{equation*}
  \frac{\Delta A(t)}{A} 
  =
  \frac{1}{A}
  \int_A \left[
    \epsilon_{xx}(\vec{r},t) + \epsilon_{yy}(\vec{r},t)
    \right] \, d^{2}\vec{r} ,
\end{equation*}
and the local rotation is
\begin{equation*}
  \Delta \theta(t)
  =
  \frac{1}{A}
  \int_A \theta_{xy}(\vec{r},t) \, d^{2}\vec{r}.
\end{equation*}
If we assume equipartition of energy, the variances in these quantities
in a box of side length $L_b$ are related to the finite-size
bulk and shear moduli $K(L_b)$ and $\mu(L_b)$ by \cite{ZahnPRL2003}
\begin{align*}
  \Var{ \frac{\Delta A(t)}{A} }_{L_b}
  & =
    \frac{k_B T}{A \left[K(L_b) + \mu(L_b)\right]}
    \quad \text{and} \\
  \Var{ \Delta \theta(t) }_{L_b}
  & =
    \frac{k_B T}{2 A \, \mu(L_b)}.
\end{align*}
The thermodynamic limits of the elastic constants are obtained by the
finite-sized scaling procedure \cite{SenguptaPRE2000} shown in
Fig.~\ref{fig:elasticConsts}(E).

The elastic moduli are related to the potential energy $U(r) = B
r^{-3}$ of the particles' pair repulsion at the nearest-neighbor
separation, $r = a$.  In terms of the dimensionless interaction
parameter $\Gamma = B (\pi \rho)^{3/2}/k_{B}T$, we expect, in the
high-density limit \cite{KeimPRL2004},
\begin{equation} 
  \Gamma  = \frac{a^{2}}{0.3461 \, k_{B}T} \, \mu \quad \text{and}
              \quad 
  \Gamma  = \frac{a^{2}}{3.461 \, k_{B}T} \, K.
\label{eq:dipoleCrystalFormulas}
\end{equation} 
Thus, each elastic constant provides a measurement of $\Gamma$. We
take the average of these measurements to be our estimate for
$\Gamma$, and half their difference to be the corresponding
uncertainty $\Delta \Gamma$.  Finally, we estimate the
nearest-neighbor interaction force $F_{r}(a) = 3 B a^{-4} = 3k_{B}T
\Gamma (\pi \rho)^{-3/2} a^{-4}$.

We confirm the accuracy of our implementation of this protocol
through molecular dynamics simulations perfomed using the HOOMD-blue
suite \cite{hoomd1,hoomd2,hoomdWebsite}.  For parameter values similar
to those of the experiment, our analysis of the particle trajectories
accurately reproduces the interaction parameter $\Gamma$ and the
interparticle force $F_{r}$.

\begin{figure}[t!]
\includegraphics[width=\columnwidth]{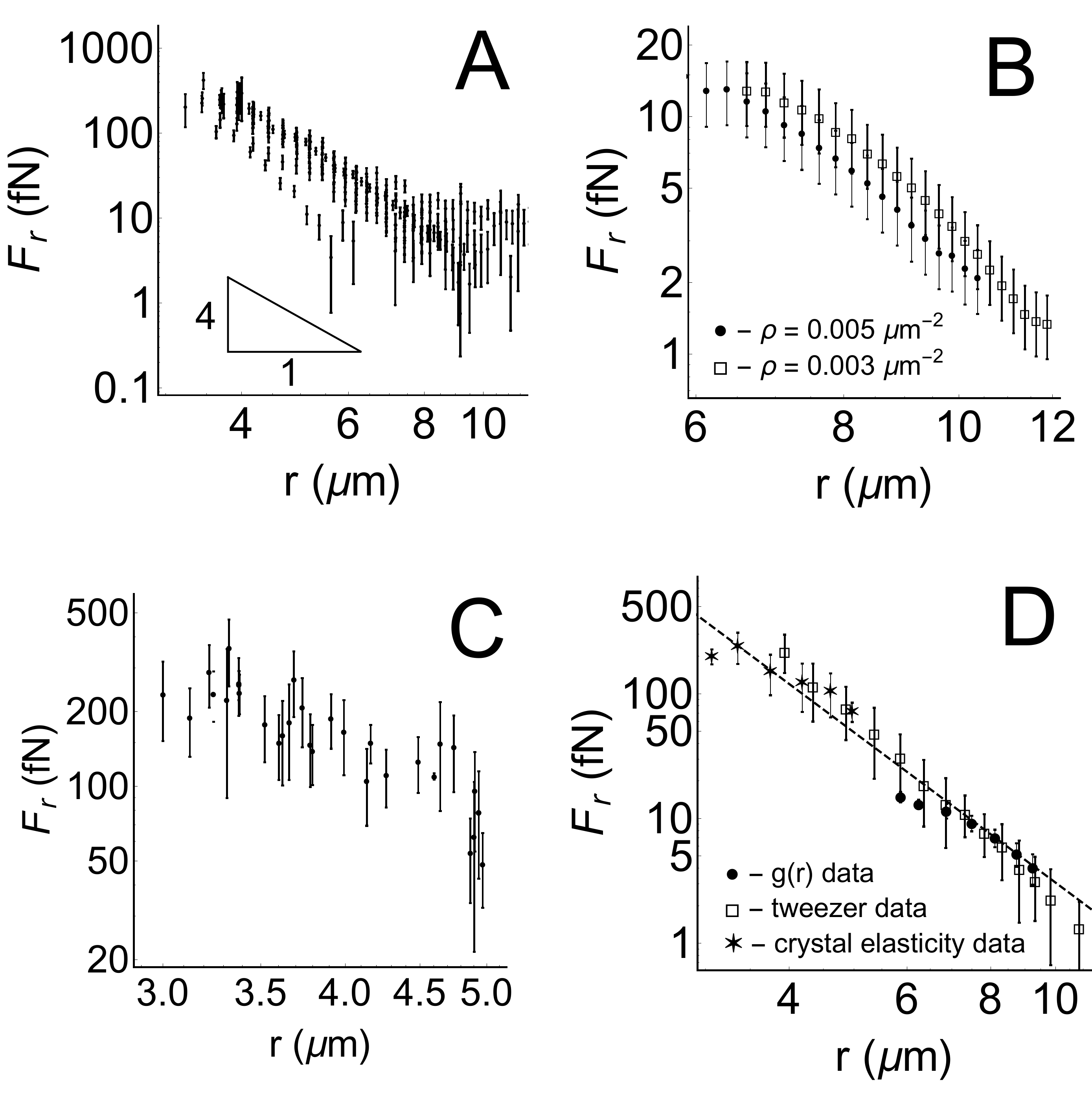}
  \caption{Log-log plots of the results of three experiments to find
forces between interfacial colloids. {{(A)}} Results of the
catch-and-release laser tweezer experiment, for 10 pairs of
interfacial particles. {{(B)}} Results of finding the interparticle
force via the pair potential $g(r)$, for two samples at different
areal densities. {{(C)}} Results of experiments where we found
the interparticle force via the elastic moduli of 2D colloidal
crystals. Each data point corresponds to a different video of a
fluctuating lattice. {{(D)}} The data from panels A, B and C are
divided into bins. For each data set, the mean in each bin is
plotted. Error bars indicate the spread of the data in a given bin:
standard deviation for data sets A and C, maximum deviation from
the mean in the case of B. The dashed line is a fit of all the data
to Eq.~\eqref{eq:Fr}, which gives $q = \SI{570(30)}{e}$.}
  \label{fig:FinalResult} 
\end{figure}

Fig.~\ref{fig:FinalResult}(C) shows the results of applying this
analysis to five different samples. We restrict
our data collection to crystals that have lattice constants between $a
= \SI{3}{\um}$ and \SI{5}{\um}.  Crystals with $a \lesssim
\SI{3}{\um}$ do not satisfy the far-field assumption $r^{2} \gg
d^{2}$, while those with $a \gtrsim \SI{5}{\um}$ do not have high
enough density to justify the use of
Eq.~\eqref{eq:dipoleCrystalFormulas}. Fitting Eq.~\eqref{eq:Fr} to the
plotted data, we find that the charge $q = \SI{590(20)}{e}$. As Table
\ref{charge_tab} shows, this value is consistent with the other pair
interaction experiments, and, within two standard deviations, is also
consistent with the results of the colloid-interface experiments.

In our experiments, we have observed the behavior of specific pairs
of particles far away from any others (tweezer experiments), as well
as systems of many particles in both the low density ($g(r)$
experiment) and the high density (crystal elasticity experiment)
limits. The fact that the measured charge is consistent in
all these cases implies that the interaction is pairwise additive over
the range of interparticle separations explored by our experiments.
This contrasts with other systems of colloidal particles dispersed in
oil \cite{Merrill2009PRL}, and may have implications for understanding
the origin of the surface charge on the particles.

\section{Conclusions}
In this work, we study the behavior of a system of charged colloids in the
vicinity of a fluid interface. We show that, in the absence of wetting
by the aqueous phase, this behavior is governed by electrostatics
alone: individual colloids interact with the interface via
image-charge attraction, while particles that are already
interfacially bound interact with their neighbors as charge-image
charge dipoles. Our model, in which the particle charge $q$ is the
only fit parameter, is consistent with data from the four independent
experiments we have performed.

In our system, interactions between interfacial particles are
pairwise-additive, are constant over time-scales of weeks, and are
homogenous enough to allow the formation of defect-free crystals over
length-scales of tens of lattice spacings. The system is thus
well-suited to the study of problems in fundamental condensed matter
physics, for example the phase behavior of repulsive particles in 2D
\cite{DeutschlanderPNAS2015, QiSM2014}, or the structure and dynamics
of topological defects in curved spaces \cite{IrvineNature2010,
IrvineNatMat2012, KusumaatmajaWalesPRL2013}. Moreover, we can now hope
to use our knowledge of electrostatic interactions in systems of
interfacial colloids to better understand the behavior of more complex
systems, such as those with partial wetting.

\vspace{0.7cm}
\begin{table}[h!]
\begin{centering}
  \begin{tabular}{ | c || c | }
    \hline
                                                                  &
                                                                    $q$
                                                                    (e)
    \\ \hline \hline
    colloid-interface experiment                 & $\quad 530 \pm 30\quad$  \\ \hline 
    laser tweezer experiment                       & $\quad 580 \pm 30 \quad$  \\ \hline
    $g(r)$ experiment                                  & $\quad 540 \pm 30 \quad$  \\ \hline
    crystal elasticity experiment                  & $\quad 590 \pm 20 \quad$ \\
    \hline
  \end{tabular}
\end{centering}
\caption{Particle charge found in each of the four experiments
  described in this paper.}
\label{charge_tab}
\end{table}

% acknowledgement.tex                            22 April 2014
%
This work was supported primarily by the National Science Foundation
under Award No. DMR-1105417.  Partial support was provided by NASA,
through Award No. NNX13AR67G, and the MRSEC program of the National
Science Foundation through Award Nos. DMR-1420073 and DMR-1420570.
G. I. G-G. acknowledges CONACYT for financial support via the program
Catedras CONACYT para Jovenes Investigadores.  The authors would like
to acknowledge fruitful discussion with Dan Evans, Gary L. Hunter,
Eric DeGiuli and Aleksandar Donev at New York University.
%
 % input acknowledgement

\appendix*
\section{Finding Strain and Rotation Tensors from Particle Trajectories}
Here we outline how we use a discretized version of the divergence
theorem from vector calculus to the calculate the strain and rotation
tensor from the particle trajectories obtained from video microscopy.

We first define the particle's equilibrium position.  After
subtracting uniform drift, we still need to eliminate effects which
are due to slow expansion, compression or rotation of the lattice, as
well as the long-wavelength fluctations which are characteristic of 2D
solids \cite{GasserChemPhysChem2010}. To do this, we use a moving
average of the particle's position using time window which is
typically 20 times the relaxation time of an individual particle, as
computed from the mean square displacement. At each frame of the
movie, the displacement $\mathbf{u}(\alpha)$ of particle $\alpha$ is
calculated relative to this moving average.  

Once we obtain the
displacement field $\mathbf{u}$ for a given frame of the movie, we
need to calculate the strain and rotation tensors, which requires
taking derivatives of $\mathbf{u}$. In order to do this, we use the
Voronoi construction to partition the field of view into cells associated with
each lattice site. This construction provides a well-defined set
of nearest neighbors for each particle, which does not change over the
course of the movies.  For an arbitrary vector field $\mathbf{v}$, the
matrix of partial derivatives $\partial_{i}v_{j}(\alpha)$ of particle
$\alpha$ can be calculated by using a discrete version of the
divergence theorem \cite{DeGiuliPRE2011}. This works as follows: for
every particle $\beta$ that neighbors $\alpha$, the particles share an
edge of a Voronoi cell, which we label ($\alpha,\beta$). For each
edge, we define the vector $\mathbf{v}(\alpha,\beta)$ as the average
of $\mathbf{v}(\alpha)$ and $\mathbf{v}(\beta)$, while the normal
vector $\mathbf{\hat{n}}(\alpha,\beta)$ and the edge length
$\ell(\alpha,\beta)$ are given by the geometry of the Voronoi
cell. Using this notation, the divergence of $\mathbf{v}$ at particle
$\alpha$ is given by
\begin{equation*} \nabla \cdot \mathbf{v}(\alpha) = \frac{1}{A(\alpha)}
\sum_{\beta \text{ n.n.}\alpha} \ell(\alpha,\beta)
\left(\mathbf{v}(\alpha,\beta) \cdot \mathbf{\hat{n}}(\alpha,\beta)
\right),
\end{equation*} where $A(\alpha)$ is the area of the Voronoi cell of
particle $\alpha$, and the sum is taken over all particles $\beta$
which are nearest neighbors to particle $\alpha$. For appropriate
choice of $\mathbf{v}$, the components of the strain and rotation
tensors can be found at each lattice site. For instance,
$\epsilon_{xx} = \partial_{x}u_{x} = \nabla \cdot \mathbf{v} $, where
$\mathbf{v} = (u_{x},0)$.

\bibliography{pre1}{} \bibliographystyle{apsrev4-1}

\end{document}